# Flocking algorithm for autonomous flying robots

Csaba Virágh[1], Gábor Vásárhelyi[1,2], Norbert Tarcai[1], Tamás Szörényi[1], Gergő Somorjai[1,2], Tamás Nepusz[1,2], Tamás Vicsek[1,2]

[1] ELTE Department of Biological Physics, 1117 Budapest, Pázmány Péter Sétány 1/A

[2] MTA-ELTE Statistical and Biological Physics Research Group, 1117 Budapest, Pázmány Péter Sétány 1/A

E-mail of corresponding author: viraghcs@hal.elte.hu

**Abstract.** Animal swarms displaying a variety of typical flocking patterns would not exist without underlying safe, optimal and stable dynamics of the individuals. The emergence of these universal patterns can be efficiently reconstructed with agent-based models. If we want to reproduce these patterns with artificial systems, such as autonomous aerial robots, agent-based models can also be used in the control algorithm of the robots. However, finding the proper algorithms and thus understanding the essential characteristics of the emergent collective behaviour of robots requires the thorough and realistic modeling of the robot and the environment as well. In this paper, first, we present an abstract mathematical model of an autonomous flying robot. The model takes into account several realistic features, such as time delay and locality of the communication, inaccuracy of the on-board sensors and inertial effects. We present two decentralized control algorithms. One is based on a simple self-propelled flocking model of animal collective motion, the other is a collective target tracking algorithm. Both algorithms contain a viscous friction-like term, which aligns the velocities of neighbouring agents parallel to each other. We show that this term can be essential for reducing the inherent instabilities of such a noisy and delayed realistic system. We discuss simulation results about the stability of the control algorithms, and perform real experiments to show the applicability of the algorithms on a group of autonomous quadcopters. Bio-inspiration works in our case two-ways. On the one hand, the whole idea of trying to build and control a swarm of robots comes from the observation that birds tend to flock to optimize their behaviour as a group. On the other hand, by using a realistic simulation framework and studying the group behaviour of autonomous robots we can learn about the major factors influencing the flights of bird flocks.

## 1. Introduction

Collective motion is an impressive phenomenon, that can be observed in a wide range of biological systems, such as fish schools, bird flocks, herds of mammals or migrating cells [1]. These systems produce the same universal feature: the velocity vectors of neighbouring individuals tend to become parallel to each other. This behaviour and the underlying control mechanism seem to be a prerequisite of safe, stable and collision-free motion. Therefore, it might be advantageous to incorporate the mathematical models that reproduce group flight patterns into the control of artificial systems, a group of autonomous flying robots, for example. By *autonomous* we mean that every agent uses on-board sensors to measure its state and performs all controlling calculations with an on-board computer, i.e., the control system is decentralized. This definition prohibits central processing of the group dynamics by an external computer, but allows the use of e.g. on-board GPS devices for external reference of position. Our study is valid for any kind of object that is capable of moving in arbitrary directions independently of its orientation, within a reasonable velocity range (including zero velocity hovering). Typical flying robots that satisfy this criteria are the so-called quadro-, hexa-, and octocopters, commonly named as multicopters.

According to Reynolds, collective motion of various kinds of entities can be interpreted as a consequence of three simple principles [2]: repulsion in short range to avoid collisions, a local interaction called *alignment rule* to align the velocity vectors of nearby units and preferably global positional constraint to keep the flock together. These rules can be interpreted in mathematical form as an agent-based model, i.e., a (discrete or continuous) dynamical system that describes the time-evolution of the velocities of each unit individually.

The simplest agent-based models of flocking describe the alignment rule as an explicit mathematical axiom: every unit aligns its velocity vector towards the average velocity vector of the units in its neighbourhood (including itself) [3]. It is possible to generalize this term by adding coupling of accelerations [4], preferred directions [5] and adaptive decision-making schemes to extend the stability for higher velocities [6]. In other (more specific) models, the alignment rule is a consequence of interaction forces [7] or velocity terms based on over-damped dynamics [8].

An important feature of the alignment rule terms in flocking models is their locality; units align their velocity towards the average velocity of other units within limited range only. In flocks of autonomous robots, the communication between the robots usually also have a finite range. In other words, the units can send messages (e.g. their positions and velocities) only to nearby other units. Another analogy between nature based flocking models and autonomous robotic systems is that both can be considered as based on agents, i.e., autonomous units subject to some system-specific rules. In the flocking models, the velocity vectors of the agents evolve individually through a dynamical system. In a group of autonomous flying robots, every robot has its own on-board computer and on-board sensors, thus the control of the dynamics is individual-based, decentralized.

Because of these similarities, some of the principles of animal flocking models can be integrated into the control dynamics of autonomous robots [9]. For example, Turgut et al. presented a dynamical system based on the simplest flocking model and used it to control the motion of so-called Kobots in two dimensions, on the ground [10]. In three dimensions, Hauert et al. presented experiments with fixed-wing agents, as a simple application of two of the three rules postulated by Reynolds (note that there was no true repulsion between the units, they flew at different altitudes) [11].

Thus the principles of flocking models presented above are useful for creating control algorithms for autonomous robots. However, we shall not underestimate the ability of animals to maintain highly coherent motion. The prerequisites of smooth collective motion include robustness against reaction times and possible delay in the communication, noisy sensory inputs or unpredictable environmental disturbances, like wind. Animals seem to overcome these difficulties efficiently. However, in robotic systems, these effects can cause unpredictable effects on stability. It is well known, for example that if time delay is present in the communication between swarming agents, instabilities can emerge [12].

One of the main goals of this paper is to provide a model of a general autonomous flying robot integrated into a realistic simulation framework. This model can be used to study the stability of flocking algorithms from the perspective of the deficiencies of realistic systems. The model should contain as many system-specific features as we can take into account, but also should be applicable for many kinds of robots. Due to this „duality" we define the axioms of the robot model with several independent parameters, corresponding to each source of deficiency in the realistic framework. Specific experimental situations can be realized with a fine-tuned set of these parameters.

Another goal of this paper is to demonstrate that some features of animal flocking models can be useful in collective robotics only if some specific extra aspects of the robots are taken into account. We show that the principles of flocking behaviour can be transformed into unique components of the dynamical system implemented as the control framework of robots. A short-range repulsion is needed to avoid collisions and an implicit viscous friction-like alignment rule term is efficiently used for damping the amplitude of oscillations caused by the imperfections of the system. With simulations and experiments on autonomous quadcopters, we study the stability of two realistic bio-inspired situations: i) a general self-propelled flocking scenario inside a bounded area and ii) a collective target tracking setup to reach and smoothly stop at a predefined position.

**2. Realistic model of a flying robot**

In this section, we present a model of a flying robot based on some features that are general in many realistic robotic systems. In such systems, the motion of the robots is controlled by a low-level algorithm, e.g. a velocity-based PID controller (see Appendix A). This low-level control algorithm typically has an input, the desired velocity vector of the actual robot. During flock flights, the time-dependence of the desired velocity of the $i$th unit can be a function of the positions ($x_i$) and velocities ($v_i$) of the other units:

$$v_i^d(t) = f_i(\{x_j(t)\}_{j=1}^N, \{v_j(t)\}_{j=1}^N),$$

where $N$ is the number of agents and the $f_i$ function contains the arbitrary features of the controlling dynamics. In ideal case, the velocity of the $i$th robot changes to $v_i^d(t)$ at time $t$ immediately. However, a robotic system is never ideal; some of its deficiencies shall be modelled:

1. **Inertia** – The robots cannot change their attitude or velocity immediately. In general, the desired velocity is an input of a low-level controller algorithm. We assume that in an optimal setup, the system can reach the desired velocity with exponential convergence, with a characteristic time $\tau_{\text{CTRL}}$. A simple controller algorithm satisfying this behaviour is a PID controller (see Appendix A). The magnitude of acceleration is also limited to $a_{\max}$.

2. **Inner noise** – We have to take into account the inaccuracy of the sensors that provide relative position and velocity information. For example, the uncertainty of the position and velocity measured by a GPS device can be modelled as a stochastic function $\eta_i^s(t)$ (see Appendix B). This function can be characterized by a standard deviation $\sigma_s$. Note that the term „inner noise" can refer to the inaccuracy of any kind of sensors used in actual robot system.

3. **Refresh rate of the sensors** – Refresh rate of sensory inputs fundamentally defines the reaction time and agility of robots. We consider a limited refresh rate of the sensors: every unit updates sensory data with frequency $t_s^{-1}$. In our current model, $t_s^{-1}$ is constant.

4. **Locality of the communication** – The communication between the units have a finite range, $r_c$, thus if the distance between two units is greater than $r_c$, they cannot interact with each other. In other words: the $f_i$ function depends on $x_j$ only if $|x_j - x_i| < r_c$.

5. **Time delay** – By the time a unit receives and processes position and velocity data from another unit, data will be old due to data processing and transmission delays. In the simplest approach, time delay can be considered as a constant value, $t_{\text{del}}$.

6. **General noise** – A delta-correlated (Gaussian) *outer noise* term $\eta_i(t)$ with standard deviation $\sigma$ is added to the acceleration of the units. This term is a model of unpredictable environmental effects such as fluctuations in the wind compensation of the low-level control algorithm.

Considering all the points above, our definition of a realistic system is the equivalent of defining the set $\{\tau_{\text{CTRL}}, a_{\max}, r_c, t_{\text{del}}, t_s, \{\eta_j(t), \eta_j^s(t)\}_{j=1}^N\}$. Time delay and communication range are hard to measure, can change randomly and have the most dangerous effects on stability. Therefore any kind of $f_i$ has to be investigated with various $t_{\text{del}}$ and $r_c$ values.

The final form of the model is an equation that defines the acceleration ($a_i(t)$) of each unit:

$$a_i(t) = \eta_i(t) + \frac{v_i^d(t) - v_i(t) - v_i^s(t)}{|v_i^d(t) - v_i(t) - v_i^s(t)|} \cdot \min\left\{\frac{v_i^d(t) - v_i(t) - v_i^s(t)}{\tau_{\text{CTRL}}}, a_{\max}\right\}, \tag{1}$$

$$\boldsymbol{v}_i^{\mathrm{d}}(t) = \boldsymbol{f}_i(\{\boldsymbol{x}_j(t-t_{\mathrm{del}}) + \boldsymbol{x}_j^{\mathrm{s}}(t-t_{\mathrm{del}})\}_{j \neq i}, \boldsymbol{x}_i(t) + \boldsymbol{x}_i^{\mathrm{s}}(t), \{\boldsymbol{v}_j(t-t_{\mathrm{del}}) + \boldsymbol{v}_j^{\mathrm{s}}(t-t_{\mathrm{del}})\}_{j \neq i}, \boldsymbol{v}_i(t) + \boldsymbol{v}_i^{\mathrm{s}}(t)),$$

where $\boldsymbol{x}_i^{\mathrm{s}}(t)$ and $\boldsymbol{v}_i^{\mathrm{s}}(t)$ give a measure of the integrated position and velocity noise for a random variable $\boldsymbol{\eta}_i^{\mathrm{s}}(t)$, which results from solving the second-order stochastic differential equation $\ddot{\boldsymbol{x}}_i^{\mathrm{s}}(t) = \dot{\boldsymbol{v}}_i^{\mathrm{s}}(t) = \boldsymbol{\eta}_i^{\mathrm{s}}(t)$. In the expression of $\boldsymbol{f}_i$, $\{...\}_{j \neq i}$ denotes a set with iterator $j \neq i$. The function $\boldsymbol{f}_i$ depends on the actual position and velocity of the $i$th agent and the delayed position and velocity of the other agents and only changes with $t_{\mathrm{s}}^{-1}$ frequency. The equations above can be solved using the *Euler* and *Euler-Maruyama* methods.

In the rest of the paper, we choose specific $\boldsymbol{f}_i$ functions with two main features:

1. $\boldsymbol{f}_i$ depends only on relative coordinates of the interacting units, i.e., no global positional information is needed in the system:

$$\boldsymbol{f}_i = \boldsymbol{f}_i(\{\boldsymbol{x}_j(t-t_{\mathrm{del}}) - \boldsymbol{x}_i(t) + \boldsymbol{x}_j^{\mathrm{s}}(t-t_{\mathrm{del}}) - \boldsymbol{x}_i^{\mathrm{s}}(t)\}_{j \neq i}, \{\boldsymbol{v}_j(t-t_{\mathrm{del}}) + \boldsymbol{v}_j^{\mathrm{s}}(t-t_{\mathrm{del}})\}_{j \neq i}, \boldsymbol{v}_i(t) + \boldsymbol{v}_i^{\mathrm{s}}(t)).$$

2. interaction terms in $\boldsymbol{f}_i$ can be expressed as a sum of local pairwise interactions ($\boldsymbol{f}_{ij}$) with other units:

$$\boldsymbol{f}_i = \sum_{j=1}^{N} \boldsymbol{f}_{ij}(\tilde{\boldsymbol{x}}_j - \tilde{\boldsymbol{x}}_i, \tilde{\boldsymbol{v}}_i, \tilde{\boldsymbol{v}}_j) \theta(r_{\mathrm{c}} - |\tilde{\boldsymbol{x}}_i - \tilde{\boldsymbol{x}}_j|),$$

where $\theta(x)$ defines the communication range explicitly; it equals to 0 if $x < 0$ and equals to 1 if $x \geq 0$. $\tilde{\boldsymbol{x}}_i$ and $\tilde{\boldsymbol{v}}_i$ are the measured position and velocity values including the modelled inner noise term: $\tilde{\boldsymbol{x}}_i = \boldsymbol{x}_i + \boldsymbol{x}_i^{\mathrm{s}}$ and $\tilde{\boldsymbol{v}}_i = \boldsymbol{v}_i + \boldsymbol{v}_i^{\mathrm{s}}$.

We also choose fixed values for some of the parameters: $t_{\mathrm{s}} = 0.2\,\mathrm{s}$, $\tau_{\mathrm{CTRL}} = 1\,\mathrm{s}$, $a_{\max} = 6\,\mathrm{m/s}^2$, $\sigma_{\mathrm{s}} = 0.005\,\mathrm{m}^2/\mathrm{s}^2$. These values represent our state-of-the-art experimental setup with quadcopters. For practical reasons, we saturate the magnitude of desired velocities expressed by the $\boldsymbol{f}_i$ functions at $v_{\max} = 4\,\mathrm{m/s}$.

In Table 1, we summarize the parameters of the model defined by (1).

*Table 1 - Parameters of the flying robot model. The column „Valid range" shows values that are valid for our experimental setup with quadcopters. For further details, see Appendix A and B.*

| Parameter | Unit | Definition | Valid range / value |
|---|---|---|---|
| $\tau_{\mathrm{CTRL}}$ | s | Relaxation time of low-level controller (e.g. PID controller) | $\tau_{\mathrm{CTRL}} \approx 1\,\mathrm{s}$ |
| $a_{\max}$ | m/s$^2$ | Maximum magnitude of acceleration | $a_{\max} = 6\,\mathrm{m/s}^2$ |
| $\sigma_{\mathrm{s}}$ | m$^2$/s$^2$ | Measure of inner noise fluctuation | $\sigma_{\mathrm{s}} = 0.005\,\mathrm{m}^2/\mathrm{s}^2$ |
| $t_{\mathrm{s}}^{-1}$ | s$^{-1}$ | Frequency of receiving sensory data | $t_{\mathrm{s}}^{-1} = 5\,\mathrm{s}^{-1}$ |
| $r_{\mathrm{c}}$ | m | Communication range | $r_{\mathrm{c}} = (30-140)\,\mathrm{m}$ |
| $t_{\mathrm{del}}$ | s | Time delay of communication | $t_{\mathrm{del}} = (0-2)\,\mathrm{s}$ |
| $\sigma$ | m$^2$/s$^3$ | Measure of outer noise fluctuation | $\sigma = (0-0.2)\,\mathrm{m}^2/\mathrm{s}^3$ |

## 3. Self-propelled flocking model

In this section, we present a minimal algorithm that is capable of driving collective robotic systems towards a stable, collision-less, self-organized correlated flocking state. This algorithm is based on the early models of animal swarms [2] [3]. By self-organization we mean that the individuals arrive at a well-defined collective state based on the units' own decisions only [13]. The desired velocity of the agents is now a sum of interaction terms and some extra terms that define the self-propelling behaviour and interactions with a bounded arena. Each term is described below in detail.

We define the agents as self-propelled particles with preferred velocity $v_{\mathrm{flock}}$:

$$\boldsymbol{v}_i^{\text{SPP}} = v_{\text{flock}} \frac{\boldsymbol{v}_i}{|\boldsymbol{v}_i|} \ . \tag{2}$$

*3.1. Short-range repulsion*

To avoid collisions, we define a local linear repulsion between the units:

$$\boldsymbol{v}_{ij}^{\text{rep}} = \frac{D(|\boldsymbol{d}_{ij}|-r_0)}{|\boldsymbol{d}_{ij}|} \boldsymbol{d}_{ij} \theta(r_0-|\boldsymbol{d}_{ij}|), \tag{3}$$

where $\boldsymbol{d}_{ij} = \boldsymbol{x}_j - \boldsymbol{x}_i$, $D$ is the strength of the repulsion, $r_0$ is the interaction range. We consider that the amplitude of fluctuations in the measured position caused by inner noise can be in the same range as $r_0$. In such a noisy system, the simple linear repulsion is superior to higher-order terms, because errors in the measured position do not cause sudden changes or singularities in the output. If the robots were able to measure their positions more accurately, higher-order terms, like the Lennard-Jones potential could be used [14].

*3.2. Velocity alignment of neighbours*

Any kind of velocity alignment rule term in realistic control algorithms should satisfy three assumptions: it should i) relax the velocity difference of units close to each other; ii) be local and iii) have an upper threshold value even when the distance between the units are close-to-zero (similar with the repulsion term). In the light of these, we implement the alignment rule with a viscous friction-like interaction term, similarly to [15] and [16]:

$$\boldsymbol{v}_{ij}^{\text{frict}} = C_{\text{frict}} \frac{\boldsymbol{v}_j - \boldsymbol{v}_i}{(\max\{r_{\min},|\boldsymbol{d}_{ij}|\})^2}, \tag{4}$$

where $C_{\text{frict}}$ is the strength of the alignment and $r_{\min}$ defines a threshold to avoid division by close-to-zero distances.

This term is a specific, practical choice for taking into account the tendency of the particles/robots to align their direction of motion. In some sense it is a discrete counterpart of the viscous friction term which would be present in a continuum description such as, e.g., the one considered first by Toner and Tu [17].

The locality of the viscous friction term in practice is guaranteed by the inverse-square decay of the term as a function of distance. However, the maximal velocity $v_{\max}$ and the value of $C_{\text{frict}}$ also has to be bounded. The interaction becomes local if the magnitude of $\boldsymbol{v}_{ij}^{\text{frict}}$ gets negligible compared to $|\boldsymbol{v}_i^{\text{SPP}}|$ at large distances, i.e. when $C_{\text{frict}} \ll v_{\text{flock}} |\boldsymbol{d}_{ij}|^2 / 2v_{\max}$ for large values of $|\boldsymbol{d}_{ij}|$. The optimal ratio of $v_{\text{flock}}$ and $C_{\text{frict}}$ is thus defined by the limit of the velocities and the desired interaction range.

*3.3. Boundaries and shill agents*

An important principle of flocking behaviour is some kind of global positional constraint that contributes to the integrity of the flock. In simulation, this feature of the positional constraint can be well substituted by using periodic boundary conditions. This is an effective method for examining the large-scale statistical properties of the system. In real experiments, periodic boundary conditions can be imitated by closing the units into a quasi-low-dimensional space, e.g. into a ring-shaped arena [18] [19], but in three dimensions these restrictions are not practical at all.

To study the flocking model with simulations, we placed the units into a square-shaped arena with repulsive walls. We define the repulsion of the wall as virtual „shill" agents [20]. If the units are outside the wall, those shill agents try to align the velocities of the units towards the centre of the arena:

$$\boldsymbol{v}_i^{\text{shill}} = C_{\text{shill}} \cdot s(|\boldsymbol{x}_a - \boldsymbol{x}_i|, \tilde{R}(\boldsymbol{x}_i, \boldsymbol{x}_a, R), d) \left( v_{\text{flock}} \frac{\boldsymbol{x}_a - \boldsymbol{x}_i}{|\boldsymbol{x}_a - \boldsymbol{x}_i|} - \boldsymbol{v}_i \right), \tag{5}$$

where $C_{\text{shill}}$ is the strength of the „shill-repulsion", $\boldsymbol{x}_a$ is the position of the centre of the arena, $s(x, R, d)$ is a sigmoid curve which smoothly reduces the strength of the repulsion inside the arena:

$$s(x, R, d) = \begin{cases} 0 & \text{if } x \in [0, R] \\ \sin\left(\frac{\pi}{d}(x-R) - \frac{\pi}{2}\right) + 1 & \text{if } x \in [R, R+d] \\ 1 & \text{if } x > R+d \end{cases}. \tag{6}$$

$\tilde{R}$ is a function that defines the shape of the arena (in this case, a square with side length $R$).

Note that the walls of the arena are pre-defined globally in the simulation, but the repulsive term only depends on the relative coordinates $\boldsymbol{x}_i - \boldsymbol{x}_a$, thus in real robotic systems the arena can be sensed locally, the same way as neighbouring units are.

The sum of the three terms defined above are the minimal prerequisites of flocking behaviour, in other words, with these terms we could guarantee stable and collision-free collective motion in our simulations and experiments:

$$\boldsymbol{v}_i^d = \boldsymbol{v}_i^{\text{SPP}} + \boldsymbol{v}_i^{\text{shill}} + \sum_{j \neq i} (\boldsymbol{v}_{ij}^{\text{rep}} + \boldsymbol{v}_{ij}^{\text{frict}}) \theta(r_c - |\boldsymbol{d}_{ij}|). \tag{7}$$

In Table 2, we summarize the parameters of the self-propelled flocking algorithm.

*Table 2 - Parameters of the self-propelled flocking algorithm*

| Parameter | Unit | Definition |
|---|---|---|
| $v_{\text{flock}}$ | m/s | Preferred „flocking" velocity |
| $D$ | 1/s | Strength of repulsion |
| $r_0$ | m | Interaction range of repulsion |
| $C_{\text{frict}}$ | m$^2$ | Strength of viscous friction |
| $r_{\text{min}}$ | m | This parameter defines a threshold to avoid division by zero |
| $R$ | m | Side length of the square-shaped arena |
| $C_{\text{shill}}$ | | Maximum strength of shill-repulsion near walls |
| $d$ | m | Characteristic „width" of the wall |

**4. Collective target tracking**

In this section, we demonstrate that the interaction terms $\boldsymbol{v}_{ij}^{\text{rep}}$ and $\boldsymbol{v}_{ij}^{\text{frict}}$ can be included in other, task-specific control algorithms. We have created a collective target tracking algorithm using an a priori defined fixed target point. The algorithm allows the units to perform a smooth transition between two stable states: the flocking state (far from the target) and the collective hovering state (around the target). During this transition near the target point, the preferred magnitude of the velocity has to approach zero smoothly and the coherence and robustness of the flock should be maintained without signs of jamming or oscillations.

Imagine the flock as a „meta-agent" at the centre of mass moving towards the target position with desired velocity $v_0$. Each unit must accomplish two tasks without collisions: i) approach this meta-agent close enough for joining the flock and ii) move parallel with the meta-agent for reaching the target collectively.

According to our definition, communication between the robots is local. Therefore, calculating the global centre of mass is physically not possible. Nevertheless, robots can calculate a *local centre of mass (CoM)*, based on the information available from within their communication range (in a sphere-shaped environment with radius $r_c$). Attraction towards the target point is thus defined as:

$$\bm{v}_i^{\mathrm{trg}} = v_0 \left[ \mathrm{s}(|\bm{x}_i^{\mathrm{CoM}} - \bm{x}_i|, r_{\mathrm{CoM}}, d) \frac{\bm{x}_i^{\mathrm{CoM}} - \bm{x}_i}{|\bm{x}_i^{\mathrm{CoM}} - \bm{x}_i|} + \mathrm{s}(|\bm{x}^{\mathrm{trg}} - \bm{x}_i^{\mathrm{CoM}}|, r_{\mathrm{trg}}, d) \frac{\bm{x}^{\mathrm{trg}} - \bm{x}_i^{\mathrm{CoM}}}{|\bm{x}^{\mathrm{trg}} - \bm{x}_i^{\mathrm{CoM}}|} \right], \quad (8)$$

where $v_0$ is the magnitude of the preferred velocity, $\bm{x}^{\mathrm{trg}}$ is the position of the target, $\bm{x}_i^{\mathrm{CoM}}$ is the position of the local centre of mass from the viewpoint of the *i*th agent, $r_{\mathrm{trg}}$ is the radius of the target area, $r_{\mathrm{CoM}}$ is the radius of the sphere-shaped meta-agent. $\mathrm{s}(x, R, d)$ is the sigmoid function defined in (6). Note that the locality of the viscous friction term defined in (4) depends on the values of $v_0$ and $C_{\mathrm{frict}}$ in this algorithm. Also note that different weights for the target and CoM tracking terms could also be introduced, but we keep these weights at 1 now to keep the algorithm as simple as possible.

The magnitude of the target tracking term saturates at $v_0$:

$$\tilde{\bm{v}}_i^{\mathrm{trg}} = \frac{\bm{v}_i^{\mathrm{trg}}}{|\bm{v}_i^{\mathrm{trg}}|} \min\{v_0, |\bm{v}_i^{\mathrm{trg}}|\}. \quad (9)$$

The final desired velocity calculated by the algorithm is:

$$\bm{v}_i^{\mathrm{d}} = \tilde{\bm{v}}_i^{\mathrm{trg}} + \sum_{j \neq i}(\bm{v}_{ij}^{\mathrm{rep}} + \bm{v}_{ij}^{\mathrm{frict}})\,\theta(r_c - |\bm{d}_{ij}|). \quad (10)$$

In Table 3, we summarize the parameters of the target tracking algorithm.

*Table 3: Parameters of the target tracking algorithm*

| Parameter | Unit | Definition |
|---|---|---|
| $v_0$ | m/s | Preferred velocity far from the target position |
| $r_{\mathrm{CoM}}$ | m | Radius of expected flock size (characteristic size of the meta-agent) |
| $r_{\mathrm{trg}}$ | m | Characteristic size of the target area |
| $d$ | m | Charasctreristic size of the „transition" area – velocity of the meta-agent approach to zero near the target point with this „relaxation length". |

**5., Results and discussion**

In this section, we present realistic simulation and robotic experiment results.

*5.1. Simulation of the flocking algorithm*

First of all, we demonstrate that typical flocking patterns can emerge even with large delays in the communication and with the presence of inner and outer noise. The coherence of the flocking state can be indicated with the order parameter

$$\psi_{\mathrm{scal}}(t) = \frac{1}{N(N-1)} \sum_{i=1}^{N} \sum_{j \neq i} \frac{\bm{v}_i(t) \cdot \bm{v}_j(t)}{|\bm{v}_i(t)||\bm{v}_j(t)|}, \quad (11)$$

where $N$ is the number of agents, $\bm{v}_i(t) \cdot \bm{v}_j(t)$ is the scalar product of two velocity vectors. In ideal flocking state, $\psi_{\mathrm{scal}} \approx 1$, in disordered state, $\psi_{\mathrm{scal}} \approx 0$. According to Figure 1, with lower $C_{\mathrm{frict}}$ values, correlated flocking behaviour with high $\psi_{\mathrm{scal}}$ cannot be observed. Higher $C_{\mathrm{frict}}$ guarantees that the emerged flocking states are stable even in the presence of noise and large time delay.

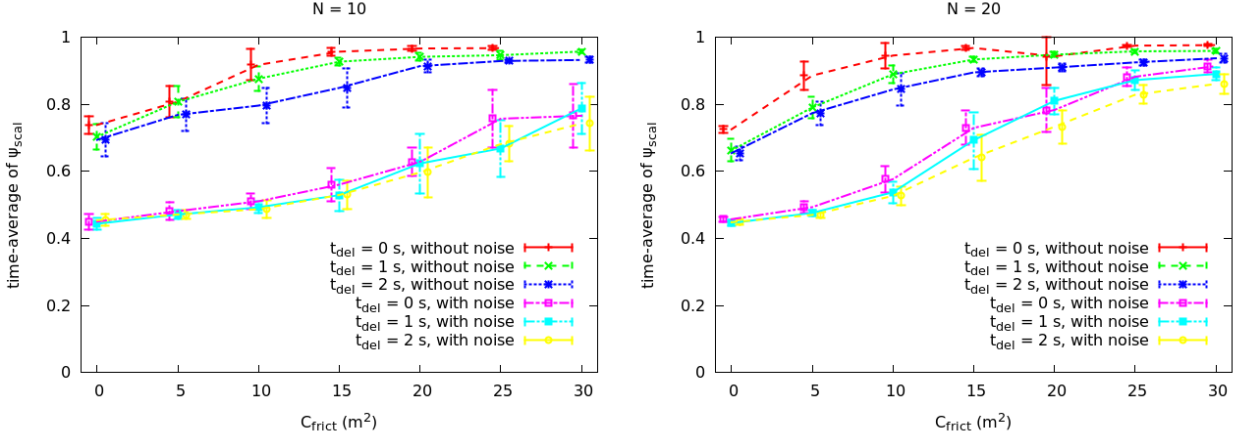

*Figure 1 - Order parameter of the flocking state ($\langle \psi_{scal} \rangle_t$) as a function of $C_{frict}$ with 10 and 20 agents. Increasing $C_{frict}$ yields more stably ordered flocks even with relatively large delay and inner and outer noises. Noise parameters are $\sigma = 0.2\,\text{m}^2/\text{s}^3$ and $\sigma_s = 0.005\,\text{m}^2/\text{s}^2$ in the simulated experiments entitled with „with noise". Other parameters: $D = 1\,\text{s}^{-1}$, $r_0 = 8\,\text{m}$, $d = 2\,\text{m}$, $r_{min} = 1\,\text{m}$, $R = 100\,\text{m}$, $C_{shill} = 2$, $r_c = 50\,\text{m}$. Every data point is averaged over 10 simulated experiments with 10 min length and different random initial conditions. Error bars show standard deviation.*

*5.2. Simulation of the target tracking algorithm*

The goal of this subsection is to show that the stability of the target tracking algorithm can be guaranteed with our selection of interaction terms used in the flocking algorithm. To study the stability, we analyze two possible quasi-stable states of the system: the flocking state (large velocity, far from the target position) and the hovering state (zero velocity, near the target position). Note that our goal is to show the effects of the interaction terms on the stability, therefore the other parameters ($v_0$, $r_{CoM}$, $r_{trg}$ and $d$) were set to fixed default values. Parameter choice was optimized to guarantee the stable completion of the target tracking task with smooth transition between the flocking and hovering states in an ideal case.

To initialize the flocking state, the units are placed within a 35 m wide square-shaped area 100 m away from the target point. After starting the simulated experiment, in ideal case, the velocity vectors of the units should become parallel and should have the same magnitude, i.e., stable, ordered flocking behaviour should be observed with $\psi_{scal} \approx 1$. We define the end of the flocking state when all units are at most $2r_{CoM}$ far from the target point. After this point, $\psi_{scal}$ shall not be used as an order parameter due to the decreased velocities around the target.

Time delay can reduce the stability of the flocking state (see Figure 2), and increasing $C_{frict}$ reduces the strength of instabilities (see right side of Figure 3).

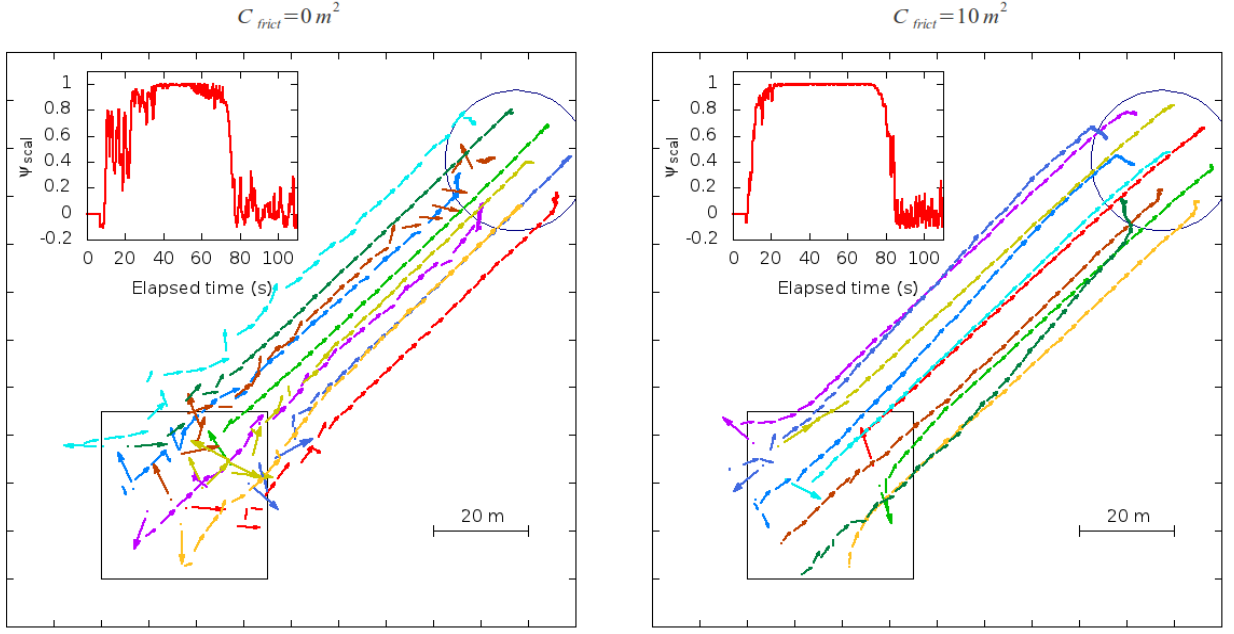

*Figure 2 - Trajectories of 10 agents with and without viscous friction, without noise, with $t_{del} = 1\,\mathrm{s}$. The arrows represent the actual velocity vector of a unit. Insets show the order parameter $\psi_{scal}$ versus time. The square represents the starting area; the circle is the environment around the target point with radius $r_{CoM} = 15.15\,\mathrm{m}$. Without the friction-like term, oscillations can emerge during the flocking behaviour, what causes severe quasi-stochastic fluctuations in $\psi_{scal}(t)$. Other parameters: $D = 1\,\mathrm{s}^{-1}$, $r_0 = 8\,\mathrm{m}$, $r_{trg} = 6.5\,\mathrm{m}$, $d = 2\,\mathrm{m}$, $v_0 = 2\,\mathrm{m/s}$, $r_c = 100\,\mathrm{m}$.*

To initialize the hovering state, units are placed around the target point inside a circle with radius $r_{trg}$ and with zero initial velocity. Due to the interaction forces and the attraction towards the target point, in ideal case, the units will arrange themselves into a lattice-like structure, where the distance between neighbours is approximately $r_0$. However, if time delay is present in the system, dangerous oscillations can emerge. Since that kind of instability can lead to collisions, it has to be eliminated. The strength of the instability in the hovering state can be described by the average velocity-magnitude:

$$\psi_{vel}(t) = \frac{1}{N} \sum_{j=1}^{N} |v_i(t)|. \quad (12)$$

Increase of $\psi_{vel}(t)$ represents growing amplitude and/or frequency of the oscillations.

In the hovering state, the strength of the delay-induced oscillations can be reduced with higher $C_{frict}$ and smaller $D$ values either with or without inner and outer noises (see left side of Figure 3). It is important to note that the instabilities can be reduced by an optimal setup of the interaction parameters only if the sphere around the local centre of mass with radius $r_{CoM}$ is large enough to contain all units with at least the repulsive interaction range ($r_0$) apart from each other. With giving the units enough space around the target, we can reduce all superfluous excitations caused by the repulsive interactions.

According to Figure 3, the additive Gaussian noise term can also reduce the instabilities caused by the time delay in the hovering state. This is a general feature of coupled delayed dynamical systems, since random noise usually acts against synchrony and resonance.

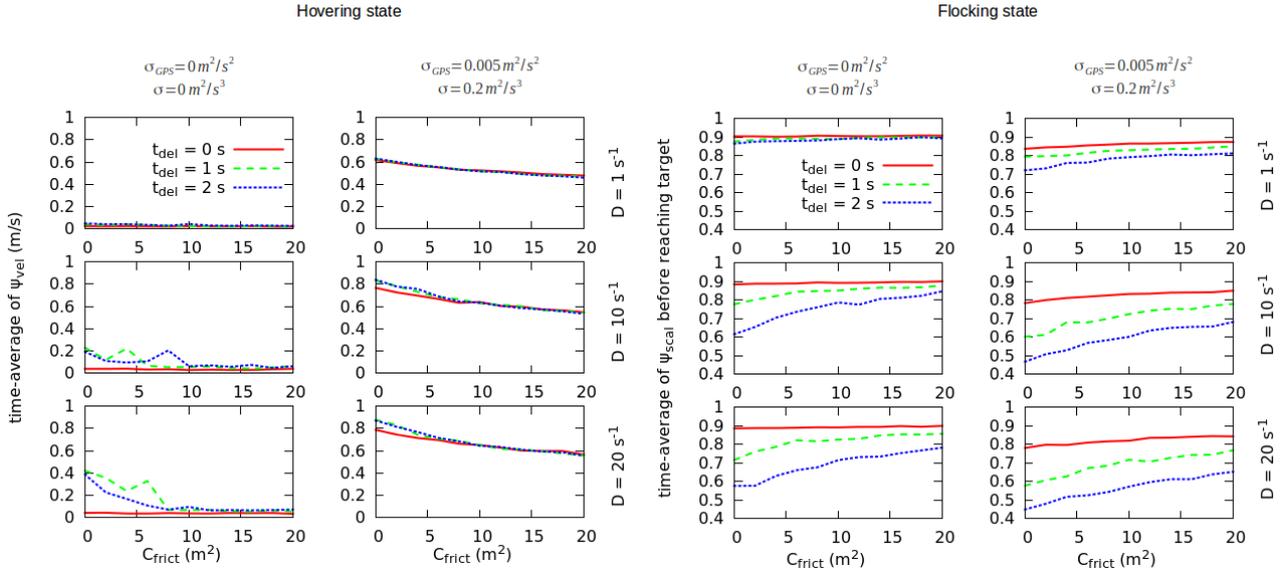

*Figure 3 - Stability of the two possible states with different $D$, $C_{frict}$ and $t_{del}$ values. Increasing $C_{frict}$ yields more stable behaviour in all cases. Other parameters: $N = 10$, $r_c = 100$m, $v_0 = 2$m/s, $d = 2$m, $r_{trg} = 6.5$m, $r_0 = 8$m, $r_{CoM}$ is $12.3$m in noiseless setups, and $13.3$m in setups with non-zero noise level. All data points are averaged over 10 simulated experiments with different random initial conditions.*

One limitation of increasing $C_{frict}$ is that it increases the overall time needed to reach the target point, especially when time delay is present in the system (see right of Figure 4). With extremely high $C_{frict}$ values, the units with zero initial velocity can practically get stuck at their initial positions.

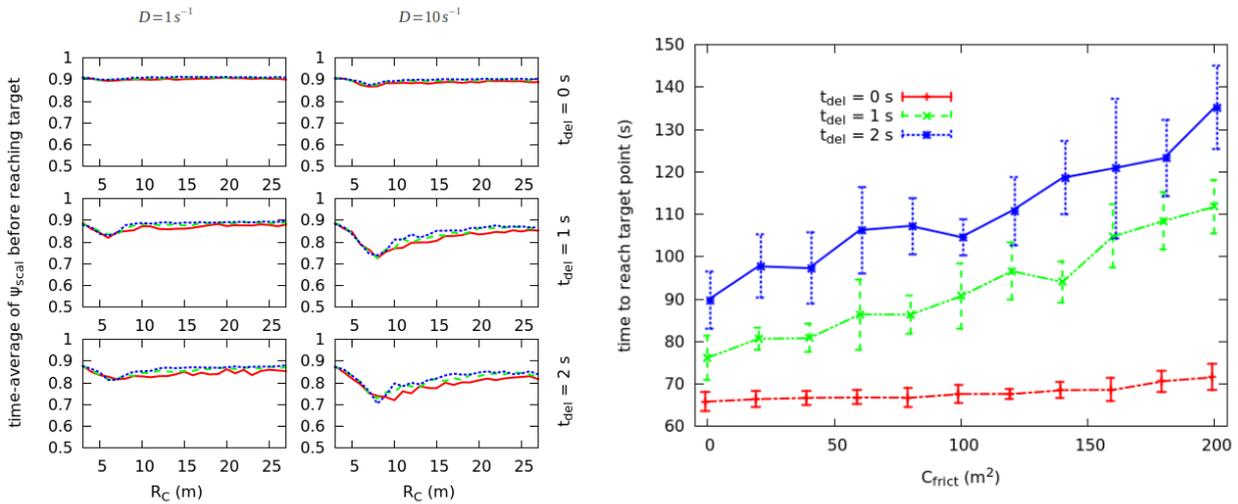

*Figure 4 - Left: time-average of $\psi_{scal}$ as a function of $r_c$ in the flocking state without noise. Increasing $C_{frict}$ increases the stability of the flocking state even with mid-range (approx. $2r_0$) $r_c$ values. Right: overall time needed to reach target point with various $t_{del}$ and $C_{frict}$ values with $r_c = 100$m and $D = 1\,\text{s}^{-1}$ without noise. Increasing $C_{frict}$ „slows down" the flock. Other parameters: $N = 10$, $v_0 = 2$m/s, $d = 2$m, $r_{trg} = 6.5$m, $r_0 = 8$m. $r_{CoM}$ is approx. $12.3$m. All data points are averaged over 10 simulated experiments with different random initial conditions. Error bars show standard deviation.*

The effects of the local communication also have to be examined. With small $r_c$ values, the units update their velocity vectors independently, thus each unit aligns its velocity towards the target point. That leads to correlated motion, but collisions can occur due to the lack of communication between the units. When $r_c$ increases close to the range of $r_0$, units can avoid collisions but they cannot organize themselves into a stable flocking state with high $\psi_{scal}$. With $r_c \gg r_0$, a correlated collective flocking state can be achieved (see left of Figure 4).

*5.3. Experiments*

We have checked the validity of the predictions of our models by implementing the algorithms presented in subsections 3.1 and 3.2 on a flying robotic flock made of 9 quadcopters. Our primary goal was to test the stability of the algorithms under realistic environmental conditions, including for example wind of a moderate level and randomly changing direction.

Our robots were based on a quadcopter (Mikrokopter L4 – ME) with an on-board Gumstix Overo Water computer. Positions and velocities were measured with U-blox Lea 6-T GPS receivers, and were sent between the robots via XBee Pro modules in broadcast mode (without establishing one-to-one connections or mesh network). Note that we used GPS for simplicity. GPS is in general not necessary for implementing the described algorithms on real robotic systems. The terms in the control algorithm depend only on the relative positions and absolute velocities. Relative coordinates were calculated using the difference of absolute positions received from GPS devices. This way, for the time being, we avoided the otherwise difficult issue of sensing position, heading and velocity of each other with local sensors [21]. The experiments were carried out outdoor, over a large plain field close to Budapest. To analyze the trajectories of the robots, we used data from GPS tracklogs. For further description of our hardware, see [22].

To test the flocking algorithm, we defined a repulsive arena as a square with 120 m sidelength around a global reference point. We performed a 20 minute measurement with 9 quadcopters moving freely inside the arena with $v_{flock}$=3.5 m/s. A two minute segment of the successful measurement is presented in the left side of Figure 5. Robots performed correlated motion while crossing the arena and changed to a new direction when they hit the wall. Some minor oscillations emerged near the walls but they always decayed quickly due to the over-damped dynamics introduced with the viscous friction term.

During the test of the target tracking algorithm, the position of the actual target point was broadcasted to the flying robots from a hand-held device in real time. We placed the target point in a car far away from the flock. After take-off, robots approached the target together and stopped above it with a smooth transition from tracking to hovering state. After some time of hovering, we drove the car over a straight trajectory and the flock followed it dynamically, still maintaining the stable, grid-like structure. Our results are presented in the right side of Figure 5.

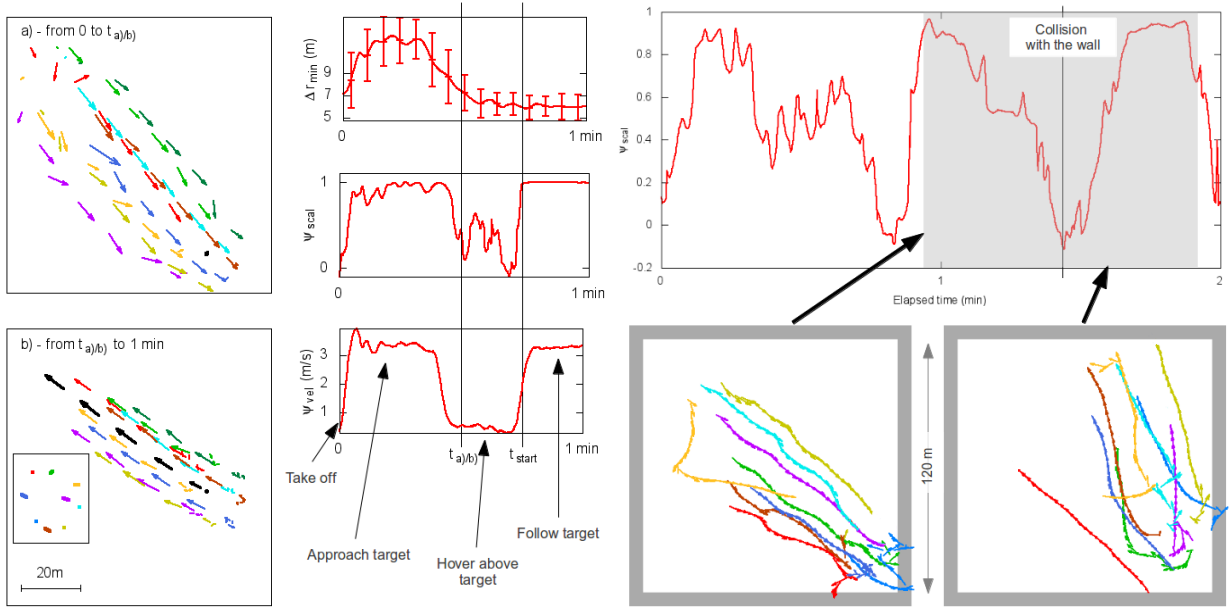

*Figure 5 – Left: Target tracking experiment with autonomous quadcopters. The boxes on the left side show the GPS trajectories of 9 robots and a target point carried by a car (thick black arrows). At the start of the measurement, the target was at a fixed position, the robots approached it with a smooth transition from flocking state to hovering state (at $t_{a)/b)}$). The inset in box b) shows the emerged grid-like pattern (robots remain static in the hovering state). At $t_{start}$, the car started to move on a straight trajectory. Next to the trajectories, three of the order parameters are shown. ($\Delta r_{min}$ is the average distance of the closest neighbours averaged over all robots). Parameters of the algorithm: $C_{frict} = 10\,m^2$, $D = 1\,s^{-1}$, $r_0 = 7\,m$, $d = 2\,m$, $v_0 = 3.5\,m/s$. Right: Experiment of the self-propelled flocking model. At the bottom of the page, trajectories of 8 agents are presented. The grey rectangles represent the arena with repulsive walls. At $t = 1.4\,min$, the flock collided with the wall, and after that, it reorganized itself into an ordered state with $\psi_{scal} \approx 1$. Parameters of the algorithm: $C_{frict} = 20\,m^2$, $D = 1\,s^{-1}$, $v_{flock} = 3.5\,m/s$, $r_0 = 10\,m$.*

We found that our flocking- and target tracking algorithm remain stable and safe even in the presence of realistic unpredictable environmental noises. For a demonstration video about some experiments, see http://hal.elte.hu/drones

*5.4 Discussion*

In this paper, we presented a realistic simulation framework for developing decentralized control algorithms for swarms of autonomous robots. This framework takes into account several realistic, but not robot-specific features, such as time delay, locality of the communication, inaccuracy and refresh rate of the sensors and inertial effects. Some of these are also present in natural swarming systems, e.g., birds. We demonstrated the applicability of this framework through the implementation of two algorithms of collective motion: a self-propelled, bio-inspired flocking algorithm and a target tracking setup. Both algorithms contain the same, carefully selected interaction terms inherited from natural flocking models: a repulsive term to avoid collisions and a viscous friction-like velocity alignment rule term for relaxing the velocities of nearby units parallel to each other.

Deficiencies of realistic systems often cause unpredictable instabilities, oscillations and collisions. With simulations, we analyzed the stability of the two algorithms, and found that the instabilities can be reduced with optimal strength of the viscous friction-like term. We implemented the optimized algorithms on a group of real autonomous robots (quadcopters with on-board computer, GPS device and XBee communication module). Successful experiments represent a direct proof of the applicability of the model and the algorithms and also show the stability of the algorithms when the system is exposed to unpredictable environmental noises.

Our realistic model can be enhanced in many ways. One big issue is synchronism vs asynchronism in the model and in reality, which appears at many levels: in the update of the simulation iterations, in the modelled delay or in the communication. We have not yet treated these issues explicitly. However, we always tested the simulation framework with larger delays compared to what was expected in our real system to overestimate the unwanted effects of the delay. Moreover, general random noise terms were introduced in order to compensate for the artificial synchrony of the used delay model. In the future we will certainly enhance our delayed communication model with asynchronous update. In the real experiments synchrony is not present at any level due to the decentralized control scheme; nevertheless, simulation results and experimental results are quite similar in general. This fact indicates that even though synchrony is an important artifact in the simulation framework, its effects are limited in the noisy environment.

In the current setup we used the generally available global positioning system as the most straightforward way of measuring position, velocity and heading. This way, we could concentrate on the development of a functional control framework in a real setup and did not have to deal with any form of "artificial vision" that is yet beyond our current knowledge. However, GPS outages could occur at any time due to several independent reasons. In the current model, GPS outages are not modelled explicitely, only through the finite sensor update rates and with the delay in the communication. In case of long periods without sensory inputs the system cannot function, per se. On the other hand, any real application requires robust behaviour. It will be an essential improvement to get around this problem when future systems become able to rely on truly local sensory information. In three dimensions this is yet an unsolved issue; however, we already designed our algorithms to be based on only local data to provide a framework for further, fully autonomous development.

Bio-inspiration was one of the main motivations of our work. Studying the analogies and differences between the behaviour of swarming robotic systems and flocking phenomena in nature reveals many important messages, some of which serve as reverse-bio-inspiration for biological research. For example, we are now inspired to search for additional factors allowing the very highly coherent motion of pigeon flocks, since our experiments suggests that a very short reaction time itself cannot account for the perfectly synchronized flight of many kinds of birds.

**Acknowledgement**

This work was supported by the FP7 ERC COLLMOT grant. No. 227878., G. V. was partly supported by EU TÁMOP 4.2.4.A/1-11-1-2012-0001

**Appendix A: PID Controller**

To model the specific features of a velocity-based PID controller, we performed measurements with real autonomous flying robots (Mikrokopter L4 – ME R/C-controllable quadcopters with a self-developed autopilot board based on a Gumstix Overo Water minicomputer). The low-level controller algorithm implemented on the on-board computer has two inputs: desired (or „target") velocity and measured velocity. The output of that controller is a control signal value fed to the standard main board of the quadrocopter. The PID loop for controlling velocity is based on the following equation:

$$\varphi_{\text{out}}(t) = K_p e(t) + K_d \frac{de(t)}{dt} + K_i \int_0^t e(t')dt' + \varphi_{\text{bias}}, \qquad (13)$$

where $e(t)$ is the *error signal*, the difference of the desired and the measured velocity: $e(t) = v^d(t) - v^m(t)$ ($v$ can be the north-south or the east-west component of the velocity vector), the $K_p$, $K_i$, $K_d$ values are the parameters of the proportional, integral and differential terms and $\varphi_{\text{bias}} = \zeta v^d$ is a feed-forward bias term determined by the linear approximation of the measured velocity as a function of the control signal. We have analyzed the logged data of our robot experiments for finding the proper parameters. The real velocity of the robot is a function of $\varphi_{out}(t)$ (See right side of Figure 6).

The time-evolution of the real velocity depends on the $K_p$, $K_i$ and $K_d$ parameters. In ideal case, an exponential convergence can be observed with characteristic time $\tau_{CTRL}$. In non-ideal case, the behaviour is either over-damped with larger settling time or under-damped with oscillations.

**Appendix B: GPS device – example for modelling inner noise**

By inner noise we mean the uncertainty of the positions and velocities measured by sensors on the robots. In the case of our quadcopters, the position data is provided by U-blox Lea 6-T GPS receivers. We have made a model to reproduce two main features of the inaccuracy of this device: i) distribution of the velocity measurement error is close to Gaussian, and ii) measured position accuracy is 2.5 m (50% CEP).

We reproduced this fluctuating behaviour of the GPS signal in an empirical model as a particle with Brownian motion in a parabolic potential centred to the real position. The Langevin equation of this situation can be expressed as the second-order stochastic differential equation $\ddot{x}^s(t) = \eta_s(t) = -D_s x^s(t) - \lambda_s \dot{x}^s(t) + \xi(t)$, where $\xi(t)$ is a delta-correlated Gaussian noise term: $\overline{\xi_i(t)\xi_j(t')} = 2\lambda_s \sigma_s \delta(t-t')\delta_{ij}$. To fit the parameters $D_s$ and $\lambda_s$, we have analyzed the fluctuating positional data of a static GPS receiver placed on the ground (for results, see left side of Figure 6). We found that with an optimal setup and the usage of the *Euler-Maruyama* method, the simulated GPS position error signal have the same characteristics as the measured data. Note that the fluctuation of position error measured with receivers placed on the ground usually have larger amplitude and frequency than the ones on a flying quadcopter, thus with this model we over-estimated the real inner noise.

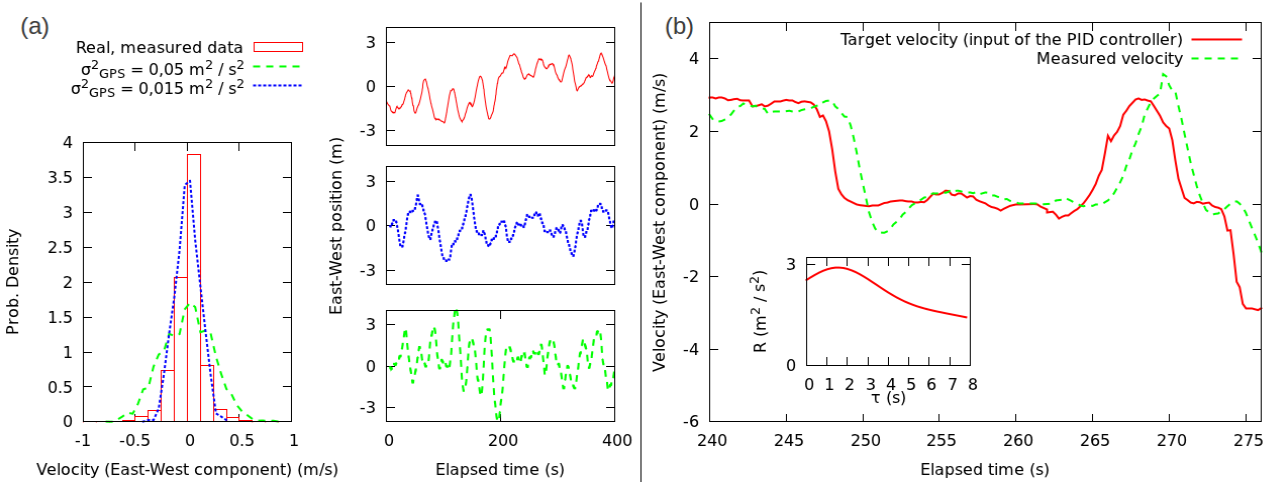

*Figure 6 - (a) Red (continuous) line: East-West position versus time (right) and velocity distribution (left) measured with a GPS device. The measurement were 20 minute long. Green and blue (dotted/dashed) lines: position as functions of time and velocity distribution modelled with different parameter settings. The parameters of the „blue/dotted" case are: $\sigma_s = 0.005 \text{m}^2/\text{s}^2$, $\lambda_s = 0.1 \text{s}^{-1}$, $D_s = \sqrt{0.2\sigma_s} (3\text{m})^{-1}$. (b) Comparison of the desired („target") velocity and the real, measured velocity of a real quadrocopter (extract from a longer measurement is shown). The inset shows the cross-correlation function of the two curves. The relaxation time of the PID controller is at the maximum of this function. In that case, $\tau_{CTRL} \approx (1-2)\text{s}$.*

**References**

[1]  T. Vicsek and A. Zafeiris, "Collective motion," *Physics Reports*, vol. 517, no. 3-4, pp. 71 – 140, 2012
[2]  C.W. Reynolds, 1987 „Flocks, herds and schools: A distributed behavioural model." *SIGGRAPH Comput. Graph*, 21, 25-34, 1987
[3]  T. Vicsek, A. Czirók, E. Ben-Jacob, I. Cohen, and O. Shochet, "Novel type of phase transition in a system of self-driven particles," *Phys. Rev. Lett.* vol. 75, no. 6, pp. 1226–1229, 1995
[4]  P. Szabó, M. Nagy, and T. Vicsek, "Transitions in a self-propelled-particles model with coupling of accelerations," *Phys. Rev. E*, vol. 79, no. 2, p. 021908, Feb 2009
[5]  J. Krause, N. R. Franks, S. A. Levin, and I. D. Couzin, "Effective leadership and decision-making in animal groups on the move," *Nature*, vol. 433, pp. 513–516, 2004
[6]  H. Dong, Y. Zhao, and S. Gao, "A fuzzy-rule-based couzin model," *Journal of Control Theory and Applications*, vol. 11, no. 2, pp. 311–315, 2013
[7]  D. Grossman, I. S. Aranson, and E. B. Jacob, "Emergence of agent swarm migration and vortex formation through inelastic collisions," *New Journal of Physics*, vol. 10, no. 2, p. 023036, 2008
[8]  B. Szabó, G. J. Szőllősi, B. Gönci, Zs, D. Selmeczi, and T. Vicsek, "Phase transition in the collective migration of tissue cells: Experiment and model," *Physical Review E*, vol. 74, no. 6, pp. 061908+, Dec. 2006
[9]  M. Brambilla, E. Ferrante, M. Birattari, and M. Dorigo, "Swarm robotics: A review from the swarm engineering perspective," *Swarm Intelligence*, vol. 7, pp. 1–41, 2013
[10] A. Turgut, H. Çelikkanat, F. Gökçe, and E. Şahin, "Self-organized flocking in mobile robot swarms," *Swarm Intelligence*, vol. 2, pp. 97–120, 2008, 10.1007/s11721-008-0016-2
[11] S. Hauert, S. Leven, M. Varga, F. Ruini, A. Cangelosi, J.-C. Zufferey, and D. Floreano, "Reynolds flocking in reality with fixed-wing robots: Communication range vs. maximum turning rate," in *Intelligent Robots and Systems (IROS), 2011 IEEE/RSJ International Conference on*, 2011, pp. 5015–5020
[12] E. Forgoston and I. B. Schwartz, "Delay-induced instabilities in self-propelling swarms," *Phys. Rev. E*, vol. 77, p. 035203, Mar 2008
[13] Floreano, D., & Mattiussi, C., "Bio-inspired artificial intelligence: theories, methods, and technologies", *MIT press*, Section 7.1, pp. 516, 2008
[14] E. Ferrante, A. E. Turgut, C. Huepe, A. Stranieri, C. Pinciroli, and M. Dorigo, "Self-organized flocking with a mobile robot swarm: a novel motion control method." *Adaptive Behaviour*, vol. 20, no. 6, pp. 460–477, 2012
[15] F. Cucker and S. Smale, „Emergent behavior in flocks." *IEEE Transactions on Automatic Control*, 852-862, 2007
[16] D. Helbing, I. Farkas, and T. Vicsek, „Simulating dynamical features of escape panic," *Nature*, 407, 487-490, 2000
[17] J. Toner and Y. Tu, "Long-range order in a two-dimensional dynamical *XY* model: How birds fly together," *Phys. Rev. Lett.*, vol. 75, pp. 4326–4329, Dec 1995
[18] C. A. Yates, R. Erban, C. Escudero, I. D. Couzin, J. Buhl, I. G. Kevrekidis, P. K. Maini, and D. J. T. Sumpter, "Inherent noise can facilitate coherence in collective swarm motion," *Proceedings of the National Academy of Sciences*, vol. 106, no. 14, pp. 5464–5469, Apr. 2009
[19] N. Tarcai, Cs. Virágh, D. Ábel, M. Nagy, P. L. Várkonyi, G. Vásárhelyi, and T. Vicsek, "Patterns, transitions and the role of leaders in the collective dynamics of a simple robotic flock," *Journal of Statistical Mechanics: Theory and Experiment*, vol. 2011, no. 04, p. P04010, 2011
[20] J. Han, M. Li, and L. Guo, „Soft control on collective behavior of a group of autonomous agents by a shill agent," *Journal of Systems Science and Complexity*, 19, 54-62, 2006
[21] C. Moeslinger, T. Schmickl, and K. Crailsheim, „A Minimalist Flocking Algorithm for Swarm Robots." *In: Kampis, G., Karsai, I., and Szathmáry, E. (Ed.), Advances in Artificial Life. Darwin Meets von Neumann*, Springer Berlin Heidelberg, 2011
[22] G. Vásárhelyi, Cs. Virágh, N. Tarcai, T. Szörényi, G. Somorjai, T. Nepusz, and T. Vicsek, „Outdoor flocking and formation flight with autonomous aerial robots." *submitted to IROS 2014,* 2014